\documentclass[format=acmsmall, review=false, nonacm]{acmart}
\usepackage{booktabs} % For formal tables
\usepackage[ruled]{algorithm2e} % For algorithms

\SetAlFnt{\small}
\SetAlCapFnt{\small}
\SetAlCapNameFnt{\small}
\SetAlCapHSkip{0pt}
\IncMargin{-\parindent}

\setcitestyle{authoryear}
\usepackage[textsize=scriptsize,textwidth=3cm,shadow]{todonotes}

\title[The Swiss Gambit]{The Swiss Gambit}

\author{\'{A}gnes Cseh}
\affiliation{
  \institution{University of Bayreuth}
  \city{Bayreuth}
  \country{Germany},
  \institution{Institute of Economics,
  Centre for Economic and Regional Studies}
  \city{Budapest}
  \country{Hungary}}
\email{agnes.cseh@uni-bayreuth.de}

\author{Pascal F\"{u}hrlich}
\affiliation{
  \institution{Potsdam Institute for Climate Impact Research}
  \city{Potsdam}
  \country{Germany}}
\email{pascal.fuehrlich@pik-potsdam.de}

\author{Pascal Lenzner}
\affiliation{
  \institution{Hasso Plattner Institute, University of Potsdam}
  \city{Potsdam}
  \country{Germany}}
\email{pascal.lenzner@hpi.de}

\begin{abstract}
In each round of a Swiss-system tournament, players of similar score are paired against each other. An intentional early loss therefore might lead to weaker opponents in later rounds and thus to a better final tournament result---a phenomenon known as the Swiss Gambit. To the best of our knowledge it is an open question whether this strategy can actually work. 

This paper provides answers based on an empirical agent-based analysis for the most prominent application area of the Swiss-system format, namely chess tournaments. We simulate realistic tournaments by employing the official FIDE pairing system for computing the player pairings in each round. We show that even though gambits are widely possible in Swiss-system chess tournaments, profiting from them requires a high degree of predictability of match results. Moreover, even if a Swiss Gambit succeeds, the obtained improvement in the final ranking is limited.
Our experiments prove that counting on a Swiss Gambit is indeed a lot more of a risky gambit than a reliable strategy to improve the final rank.
\end{abstract}

\begin{document}

\maketitle

\section{Introduction}

At sports tournaments, players and teams are usually expected to give their best in each game they play. However, they might withhold effort for strategic reasons: for example, a team might let weaker team members play a match they would probably lose anyway, so their strong team members are well rested when playing a more important match. A considerably more obscure scenario happened at the Olympic Games 2012: two badminton teams were both trying to lose a group-stage match, because the loser would meet a weaker team at the upcoming elimination match \cite{greene2012olympic}.

\subsection{Gaming the Swiss System}\label{sec:swiss_gambit}
The \emph{Swiss-system} tournament format is widely used in competitive games like most e-sports, badminton, and chess, the last of which this paper focuses on. In such tournaments, the number of rounds is predefined, while the pairing of players in each of these rounds depends on the results of previous rounds. In each round, players of similar score are paired against each other. Therefore, a weaker performance generally leads to weaker opponents.

These rules provide an incentive to intentionally draw or even lose a match early in a tournament to subsequently play against weaker opponents and finish the tournament at a better rank than if that early match would have been won. This strategic move is colloquially called a \textit{Swiss Gambit}, referring to a gambit in chess, which means sacrificing a piece in order to gain an advantage~\cite{wikipedia2019gambit}. However, even though some observations based on intuition or minimalistic simulations, such as ``this strategy [\ldots] is just as likely to backfire as succeed''~\cite{Ead16} has been made, to the best of our knowledge, no research study has been conducted on the Swiss Gambit, despite the fact that it is well-known among competitive chess players~\cite{Ead16,Gen21,Gua21}, and the lack of ``serious research'' has been pointed out in their online discussions~\cite{Quo21,Red21,Sta21}.

Intentionally losing a match to gain an advantage in a tournament is a highly controversial strategy, which is generally considered unethical or even cheating. In 2019, the five-time ex-world champion Vishy Anand has been tagged by punters, pundits, and commentators alike as having ``played the gambit'', when staging a remarkable recovery following his shock opening-round upset result at the Grand Swiss tournament on the Isle of Man~\cite{Hen21}.

\subsection{Related Literature}
The Swiss system regularly evokes interest in the AI and Economics communities~\citep{Van13,HAH16,BXH+18, LG22,SBC22,FCL22}. The works of \citet{Csato13,Csato17,Csato21} study the ranking quality of real-world Swiss-system tournaments, in particular, whether a fairer ranking could have been obtained by different scoring rules. The importance of winning or losing compared to drawing a lot of games was highlighted by~\citet{Bil06}. Computing player pairings at Swiss-system chess tournaments is also a popular topic. Automated matching approaches are proposed by \citet{glickman2005adaptive}, while \citet{kujansuu1999stable} use the stable roommates problem, see \cite{irving1985efficient}, to model a Swiss-system tournament pairing decision. \citet{olafsson1990weighted} and \citet{BFP17} attempt to implement the official FIDE criteria as accurately as possible. \citet{FCL22} propose a new approach to derive fair pairings at tournaments and they analyze the obtained ranking quality.

Sports tournaments are by far not the only application area of the Swiss system. Self-organizing systems~\cite{FW09}, person identification using AI methods~\cite{WTW+15}, and choosing the best-fitting head-related transfer functions for a natural auditory perception in virtual reality~\cite{OVF19} all rely on the Swiss system as a solution concept.

\subsection{Structure of the Paper and Our Contribution}

In Section~\ref{sec:background}, we describe the Swiss system used at official chess tournaments organized by the International Chess Federation (FIDE). Then, in Section~\ref{sec:models} we introduce our two models to capture a tournament. Gambit heuristics to exploit these models are then discussed in Section~\ref{sec:identifying_gambit_possibilities}. We introduce how we measure the impact of gambits in Section~\ref{sec:measuring_the_impact} and describe our simulation settings in Section~\ref{sec:gambit_simulation_settings}. Finally, the results of our simulations are presented in Section~\ref{sec:gambit_simulation_results}.

Our key insights on the impact of gambits are:
\begin{enumerate}
    \item Gambits are possible even in very small tournaments (Section~\ref{sec:background}).
    \item There is an effective gambit heuristic even if
    match results can only be approximated (Section~\ref{sec:gambit_heuristics_in_the_probabilistic_model}).
    \item The gambit player must be able to estimate match results very accurately in order to identify a gambit possibility (Section~\ref{sec:number_of_gambit_possibilities}).
    \item Even with a successful gambit, the expected rank improvement is small. (Section~\ref{sec:mean_rank_difference}).
    \item Gambits are more likely to succeed if the players' strengths span a large range or if the tournament is long (Section~\ref{sec:total_rank_difference}).
    \item The impact of gambits on the ranking quality of all players is low in general (Section~\ref{sec:impact_of_gambits_on_ranking_quality}).
\end{enumerate}
 All in all, our research establishes that even though the Swiss Gambit might lead to a higher rank in tournaments, under realistic conditions it cannot be used to reliably improve a player's rank or to derive a a largely false final ranking.

\section{The Swiss System in Chess}\label{sec:background}

\textit{Players} are entities participating in a Swiss-system tournament. Each player has an Elo rating, which is a measure designed to capture her current playing \emph{strength} from the outcome of her earlier matches \cite{elo1978rating}. In a \textit{match} two players, $a$ and $b$, play against each other. The three possible \textit{match results} are: $a$ wins and $b$ loses, $a$ and $b$ draw, $a$ loses and $b$ wins. The winner receives 1 point, the loser 0 points, while a draw is worth 0.5 points. A Swiss-system tournament consists of multiple \textit{rounds}, each of which is defined by a \textit{pairing}: a set of disjoint pairs of players, where each pair plays a match. At the end of the tournament, a strict ranking of the players is derived from the match results.

\paragraph*{Computing the Pairing in Each Round}
\label{sec:bbp_engine}
A \textit{pairing engine} calculates the pairing of players for each round, based on the results of previous rounds. The pairing must adhere to specific rules, such as no two players play against each other more than once in the tournament, the number of games played with black and white pieces is balanced for each player, and opponents playing a match should have equal or similar score.

\begin{figure}[ht]
\centering
\includegraphics[width=0.6\linewidth]{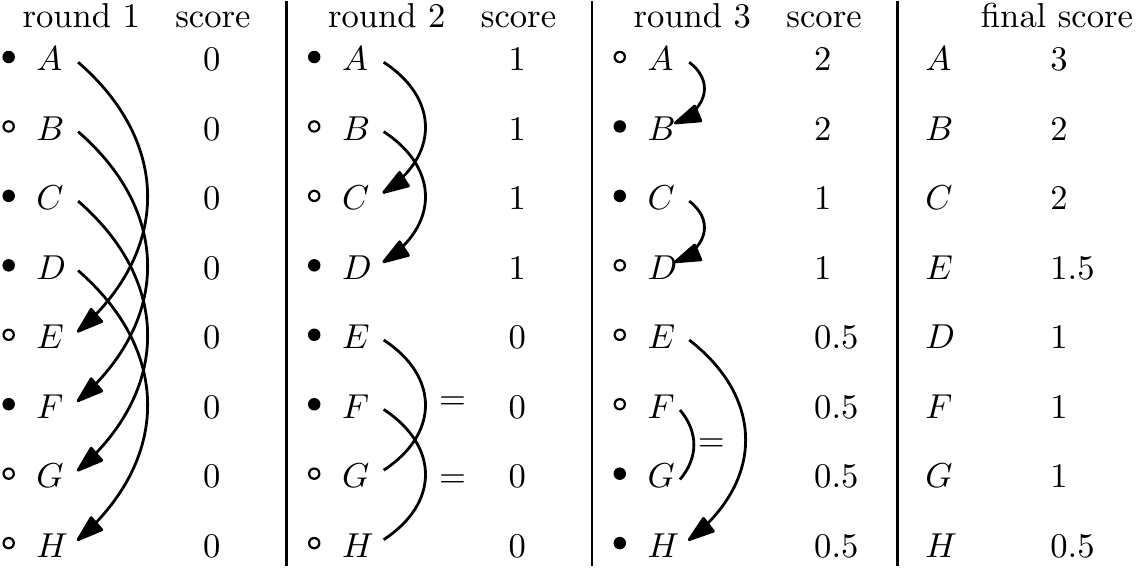}
\caption{Example of a 3-round Swiss-system chess tournament paired with the Dutch pairing system. The players $A,B,C,D,E,F,G,H$ are labeled according to their strength, with player $A$ being the strongest. Arcs indicate the matches of the respective rounds. Arrows point from winner to loser while undirected arcs indicate a draw. The initial score and the color distribution ($\bullet$ for black and $\circ$ for white) in a round are shown in each column. The final ranking and final scores are shown on the right. All players play truthfully.}
\label{fig:no_gambit}
\end{figure}

\begin{figure}[ht]
\centering
\includegraphics[width=0.6\linewidth]{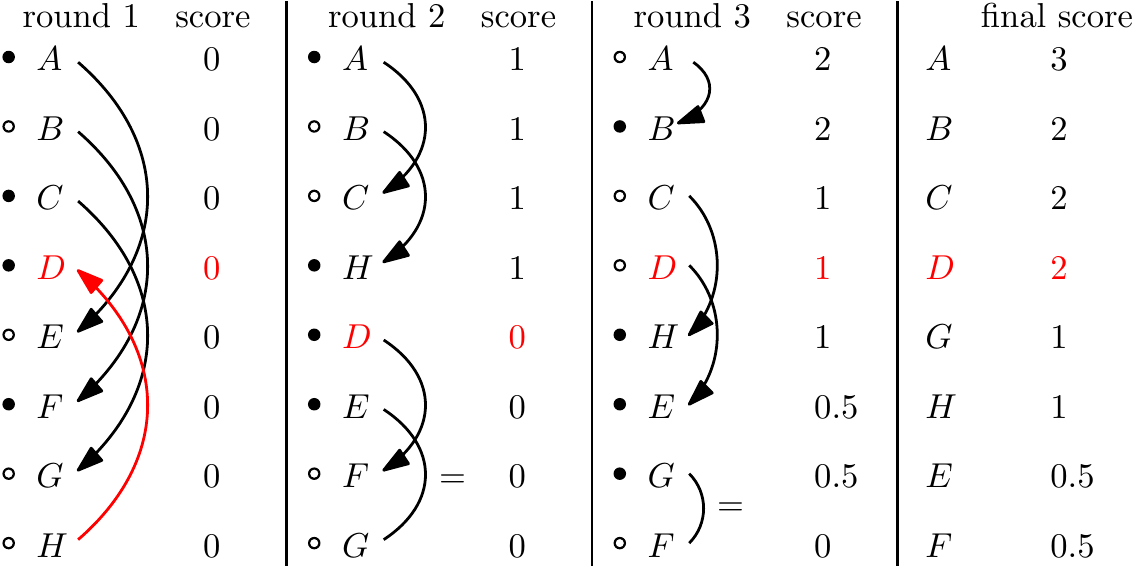}
\caption{The tournament from Figure~\ref{fig:no_gambit} with player $D$ intentionally losing her first match. This is a successful Swiss Gambit since player $D$ gains one rank and one point compared to playing truthfully as in Figure~\ref{fig:no_gambit}.}
\label{fig:gambit}
\end{figure}

The open-source and state-of-the-art pairing engine \emph{Dutch BBP}, developed by
\citet{bierema2017bbp}, is endorsed by the FIDE~\cite[C.04.A.10. Annex-3]{fide2020handbook}.
It implements the voluminous FIDE pairing criteria strictly \cite[C.04.3 and C.04.4.2]{fide2020handbook} for the so-called Dutch pairing system and outputs the unique pairing adhering to them.
In all our experiments we use Dutch BBP for pairing the players in each round. Readers interested in the details of the Dutch pairing system can consult Appendix~\ref{app:dutch}. See Figure~\ref{fig:no_gambit} for an illustration of an example Swiss system chess tournament with its respective pairings for each round. This example is completed by the example of a successful Swiss Gambit depicted in Figure~\ref{fig:gambit}.

\paragraph*{Computing the Final Ranking}
The major organizing principle for the final ranking of players is obviously the final score. Players with the same final score are sorted by tiebreakers. The FIDE \cite[Chapter C.02.13]{fide2020handbook} defines 14 types of tiebreakers, and the tournament organizer lists some of them to be used at the specific tournament. If all tiebreaks fail, the tie is required to be broken by drawing of lots. The tiebreakers we use for obtaining the final tournament ranking are based on the current FIDE recommendation \cite[C.02.13.16.5]{fide2020handbook}.

\section{Our Models}
\label{sec:models}

We work on two models, a deterministic and a probabilistic one. In the latter more sophisticated model, the results of the individual matches are computed via a probabilistic calculation that is designed to be as realistic as possible. Due to the probabilistic setting, starting with the same set of players, the final ranking might be different in different runs. This is not possible in the deterministic model, where match results can be reliably calculated from the players' Elo rating and their color assignment.

\subsection{Probabilistic Model}\label{sec:probabilistic_model}

In the probabilistic model, match results are drawn at random from a suitably chosen probability distribution based on the players' Elo rating and on the assigned colors for the respective matches. In order to be able to derive the most realistic results, we use the probability distribution proposed by %Milvang~
\citet{milvang2016prob}, which was featured in a recent news article of the FIDE commission System of Pairings and Programs~\cite{fide2020news}. Milvang's probability distribution was engineered via a Data Science approach that used real-world data from almost 4 million real chess matches from 50\,000 tournaments.

According to Milvang's approach, the probability of a certain match outcome depends on the Elo rating of the involved players. Draw probability increases with mean Elo rating of the players. The probabilities also depend on colors, as the player playing with white pieces has an advantage. See Table~\ref{tab:example_probabilities} for some example values drawn from Milvang's distribution. 
\begin{table}[ht]
\centering
\begin{tabular}{rccc}
Player Elo ratings   & Win White  & Win Black & Draw\\
1200 (w) vs 1400 (b) & 26 \%& 57 \%& 17 \%\\
2200 (w) vs 2400 (b) & 14 \%& 55 \%&31 \%\\
2400 (w) vs 2200 (b) & 63 \%& 11 \%& 26 \% 
\end{tabular}
\caption{Example match outcome probabilities drawn from Milvang's probability distribution \cite{milvang2016prob}.
}
\label{tab:example_probabilities}
\end{table}

\subsection{Deterministic Model}\label{sec:deterministic_model}
The deterministic model is a variant of the probabilistic model: the same probabilities are calculated, but instead of drawing match results randomly, the match ends in a draw if the probability for a draw is at least 20\%, otherwise the player with higher Elo rating wins. The threshold of 20\% was chosen so that for a strength range size of 800 the relative number of draws in the probabilistic and deterministic model is equal.

\section{Identifying Gambit Possibilities}\label{sec:identifying_gambit_possibilities}

As usual in game theory, we assume that all players but one, who is called the \textit{gambit player}, are trying their best to win each game they play. Furthermore, the gambit player performs exactly one gambit.
We assume that the decision whether to perform a gambit or not, is made after all other matches of that round are finished. In a real-world tournament, the gambit player could deliberately prolong her match until all other matches are finished, so she can make the gambit decision based on more information.

A \textit{course of a tournament}, just called a \textit{course} in the following, includes the pairings and match results of the rounds that have already been paired/played. For a \textit{prefix} course, some rounds are still left to be paired/played, while a \textit{complete} course includes the pairings and match results of all rounds. A player's \textit{expected final rank} is the expected value of her final rank, given the prefix course of the tournament that was already played. The expected final rank of player $p$ is calculated as follows. \[
\mathbb{E}\left(\text{rank}(p, c')\right) = \sum_{c \in C(c')} \text{probability}(c) \cdot \text{rank}(p, c)\]
In the formula, $c'$ is a prefix course which already includes player $p$'s match result for the current round, $C(c')$ is the set of all possible completed courses given the prefix course $c'$, and rank$(p, c)$ is the rank of player $p$ at the end of the completed course~$c$. In words, the expected final rank after a decision to gamble or not is calculated as the probability of each completed course times the final rank of $p$ at the end of this completed course, summed up over all completed courses that are still possible after the decision.

Naturally, the gambit player is required to have multiple so-called \textit{match result options} to choose from: if she would normally win a match, then she can choose to win, draw or lose the match. If a match would normally end in a draw, we assume that the gambit player can choose between draw and loss.
The match result without gambit is called the \textit{actual match result}, while other match result options are called \textit{gambit match result options}. The goal of a \textit{gambit heuristic} is to select the match result option that has the best expected final rank. If this differs from the actual match result option, then we say that the gambit is \textit{beneficial}.
A prefix course is a \textit{gambit possibility} for a player, if she could improve her expected final rank by performing a Swiss Gambit in her match in the current round according to a gambit heuristic.

\subsection{Gambit Heuristic in the Deterministic Model}\label{sec:gambit_heuristic_in_the_deterministic_model}
In the deterministic model, see Section~\ref{sec:deterministic_model}, a gambit possibility can be identified as follows: for each match result option, simulate the rest of the tournament once. Choose the match result option which led to the best rank. Since the model is deterministic, the result from the simulation will be the same as the actual final result. For each match result option, only one tournament has to be simulated, so the computational cost is very low. This gambit heuristic is optimal, because it always identifies the match result option with the best expected final rank correctly.

\subsection{Gambit Heuristic in the Probabilistic Model}\label{sec:gambit_heuristics_in_the_probabilistic_model}
Identifying gambit possibilities in the probabilistic model, see Section~\ref{sec:probabilistic_model}, is far more difficult compared to the deterministic model. The na\"ive approach would be to calculate the expected final rank for each match result option. However, the number of possible courses grows exponentially with the number of players and rounds, which makes it infeasible to calculate the probabilities and ranks for all possible courses.

Instead, we sample the set of all possible courses of the given tournament by simulating the tournament several times. Consider a player $p$ winning a match in round~$x$. In this situation $p$ could intentionally lose instead of winning. We first simulate the rest of the tournament $n$ times assuming that $p$ wins in round $x$ and we record $p$'s final rank each time. This array of final ranks is the sample where ``win'' is the actual match result. Then we generate the sample where ``lose'' is the gambit match result in the exact same way, except that $p$ now loses in round~$x$.

Based on these two samples, a gambit decision can be made in a variety of ways. We call these ways gambit heuristics. One could e.g.\ compare the means or medians of the samples. In our simulations, the \textit{$p$-value heuristic} proved to be the most successful. This heuristic only suggests to do the gambit if the final ranks in the gambit sample are significantly better than in the actual match result sample.

Using the samples for the gambit match result option and the actual match result, we calculate a $p$-value using Welch's one-tailed $t$-test %\plcom{Citation?} 
and reject the null hypothesis ``a gambit does not improve the expected final rank of the respective player'' if $p < 0.05$. Our samples do not have equal variances, so we use Welch's $t$-test~\cite{MS92}, which is an adaptation of Student's $t$-test~\cite{Stu08} intended for two samples having possibly unequal variances. Both tests measure the mean equality for two groups.
%unlike Student's $t$-test \plcom{Citation?} it does not assume equal sample variances.\accompl{Should we cite sg / explain these?}
%\plcom{Yes, we should cite and give a high-level explanation.}\accompl{I added a very short explanation and asked a friend to point me to some standard statistic book.}

\paragraph{Further Gambit Heuristics in the Probabilistic Model}

The expected final rank in the probabilistic model can be approximated from the computed samples, e.g., by calculating the mean or the median of the final ranks of the samples. A larger set of samples allows more precise approximation, but it also requires more computational power. The corresponding gambit heuristic is called \textit{mean heuristic} or \textit{median heuristic}, depending on the aggregation.

Another gambit heuristic is based on expected values as match results: for each match result, simulate the rest of the tournament once and record the gambit player's final rank. However, instead of randomly drawing match results from a probability distribution based on the players' Elo rating, both players get fractional points according to their expected values, which are calculated from their Elo rating. Assume player $a$ plays a match against player $b$, and $a$ is slightly stronger than $b$. Then $a$ might have a score of 0.52 while $b$ might have a score of 0.48 after the match. This gambit heuristic is called \textit{expected value heuristic} and it returns the match result option which led to the best final rank. In contrast to the previous approach, this approach requires very little computational power, since only one tournament simulation is needed for each match result option.

We implemented and simulated both of the above described gambit heuristics for the probabilistic model, but results were not convincing. In contrast to the $p$-value heuristic from Section~\ref{sec:gambit_heuristics_in_the_probabilistic_model}, gambit players actually lost ranks on average when using one of the other heuristics.

\section{Measuring the Impact of Gambits}
\label{sec:measuring_the_impact}

To quantify the impact of gambits on a specific tournament, we consider four measures: the number of gambit possibilities, the mean rank difference, the total rank difference, and the ranking quality. In this section we define and illustrate these measures using the example from Figures~\ref{fig:no_gambit} and~\ref{fig:gambit}.
\subsection{Number of Gambit Possibilities}
The \textit{number of gambit possibilities} is the total number of times the employed gambit heuristic came to the conclusion that a gambit is beneficial. Naturally, a high number of gambit possibilities means that the system can often be gamed.

One gambit possibility is shown in Figure~\ref{fig:gambit}. In the deterministic model, where match results can be predicted, the tournament in Figure~\ref{fig:no_gambit} offers no further gambit possibility.

\subsection{Rank Difference}
The \textit{rank difference} of a gambit possibility is the gambit player's final rank in the simulation with a gambit minus her rank in the simulation without a gambit. A rank difference of -2 means that the gambit player improves by two ranks, e.g., she finishes at rank 4 without a gambit and ends up at rank 2 with a gambit.

A tournament's \textit{mean rank difference} is the mean of all the rank differences of all gambit possibilities. A mean rank difference of -1.4 means that in an average gambit possibility, the gambit player will improve by 1.4 ranks. However, since the number of gambit possibilities is not taken into account, this measure can be misleading: in a tournament with a single gambit possibility with rank difference -10, the mean rank difference will also be -10 for the whole tournament, even though gambits are virtually never beneficial.

As the only gambit possibility leads to the gambit player improving her rank by 1 in the example in Figure~\ref{fig:gambit}, the mean rank difference is~-1.

A tournament's \textit{total rank difference} is the sum of all the rank differences of all gambit possibilities. Strongly negative values indicate that players are strongly incentivized to perform a gambit. The total rank difference is equal to the number of gambit possibilities multiplied by the mean rank difference, so it provides the best big-picture indicator for how much players are incentivized to gambit in a given tournament.

As the only gambit possibility leads to the gambit player improving her rank by 1 in the example in Figure~\ref{fig:gambit}, the total rank difference is~-1.

The mean rank difference can be misleading at times, for example when very few gambits are possible. However, on other classes of instances also the total rank difference can be misleading, for example if many insignificant improvements can be made. Hence, we analyze both measures. As we will see, the obtained insights are indeed similar, which proves that even though both metrics can be misleading, our results are not due to those misleading instances.

\subsection{Ranking Quality}
The ranking quality measures how similar the tournament's final ranking is to the \emph{ground-truth ranking}, which sorts the players by their Elo rating. The most popular measure for the similarity of two rankings is presumably the Kendall $\tau$ distance \cite{kendall1945treatment}. It counts the number of discordant pairs: these are pairs of elements $x$ and $y$, where $x < y$ in one ranking, but $y < x$ in the other ranking. We use its normalized variant, where $\tau \in [-1, 1]$, and $\tau = 1$ means that the rankings are identical, while $\tau = -1$ means that one ranking is the inverse of the other ranking. A higher Kendall $\tau$ distance is better, because it indicates a larger degree of similarity between the ground-truth and the output ranking.

A gambit possibility's \textit{Kendall $\tau$ difference} is the difference of two Kendall $\tau$ distances from the ground-truth ranking. It is calculated by subtracting the Kendall~$\tau$ distance of the final ranking without the gambit and the ground-truth ranking from the Kendall~$\tau$ distance of the final ranking with the gambit and the ground-truth ranking. A positive Kendall~$\tau$ difference means that the ranking with gambit is closer to the ground-truth ranking than the obtained ranking without gambit. The \textit{mean Kendall~$\tau$ difference} is the mean of the Kendall~$\tau$ differences of all gambit possibilities. It indicates how much the ranking quality is changed by gambits.

A mean Kendall~$\tau$ difference of zero means that gambits have no effect on the ranking quality. A positive mean Kendall~$\tau$ difference means that gambits improve ranking quality, which indicates a poor ranking quality of the tournament format in general. A negative mean Kendall~$\tau$ difference would be expected, because the gambit player misuses her actual Elo rating. However, the rank of the gambit player has only a small effect on ranking quality, so a strongly negative mean Kendall~$\tau$ difference mostly indicates that the non-gambit players are ranked poorly due to the gambit.

The final ranking in Figure~\ref{fig:no_gambit} contains only one discordant pair, and therefore is of normalized Kendall~$\tau$ distance $1 -\frac{2\cdot 1}{21}$ from the ground-truth ranking. The final ranking with the gambit, depicted in Figure~\ref{fig:gambit}, contains four discordant pairs, and is of normalized Kendall~$\tau$ distance of $1 - \frac{2\cdot 4}{21}$ from the ground-truth ranking. Therefore, the Kendall~$\tau$ difference of the gambit possibility is $\left(1 -\frac{2\cdot 4}{21}  \right) - \left(1 -\frac{2\cdot 1}{21}  \right) = -\frac{2}{7}$. As the tournament only admits one gambit possibility in total, it is also the mean Kendall~$\tau$ difference.

\section{Experimental Setup}\label{sec:gambit_simulation_settings}
We present the details of our agent-based simulations.
\subsection{Simulation Parameters} We ran our simulations with the optimal gambit heuristic described in Section~\ref{sec:gambit_heuristic_in_the_deterministic_model} for the deterministic model, and with the $p$-value heuristic from Section~\ref{sec:gambit_heuristics_in_the_probabilistic_model} for the probabilistic model.
The following parameters were used, unless stated otherwise:
\begin{itemize}
    \setlength\itemsep{0mm}
    \item number of players: 32
    \item number of rounds: 5
    \item number of tournaments: 1000
    \item probabilistic model sample size: 200
    \item strength range: 1000--2600
\end{itemize}

Most non-professional tournaments take place on one day only and the pool of players is rather diverse. The above values were chosen to be as realistic as possible for such an event, based on parameters of more than 320\,000 real-world tournaments uploaded to the website \url{chess-results.com}.\footnote{The data was kindly provided by Heinz Herzog, author of the FIDE-endorsed tournament manager \url{Swiss-Manager} \citep{herzog2020swiss} and \url{chess-results.com} \citep{herzog2020chess}.} To draw a complete picture, we also tested our models on tournaments with more rounds and a smaller strength range for comparison---see the plots in Section~\ref{sec:gambit_simulation_results}. 

Players were sampled uniformly at random from the strength range. For each given tournament, we first used the Dutch BBP engine to pair the players as the rules of the FIDE prescribe. The results of each match were then calculated according to the rules of the deterministic or the probabilistic model. The lead us to the final ranking. We remind the reader that the final ranking might be different for different runs of the same tournament in the probabilistic model.

\subsection{Computational Load} 

In order to identify all gambit possibilities and measure their effect, we reran the tournament several times, always assuming that a gambit is performed in a chosen match. If the match has a winner, then this player has three match result options: win, draw, and lose, which results in 600 tournament simulations in total. For a match that ends in a draw, 800 tournament simulations are required, as both players can decide between draw and loss. The prefix course consists of all other matches in the current round and all matches in previous rounds.

We forwent calculating gambit heuristics in the last two rounds, because gambits are never beneficial in the last round and almost never in the second to last round. Therefore, a tournament with 32 players and 5 rounds consists of a total of $\frac{32}{2} \cdot (5-2) = 48$ matches in which the gambit heuristic is calculated. This means that for a single complete tournament simulation with all gambit possibilities, at least $48 \cdot 600 = 28\,800$ prefix tournament simulations are needed. However, these are not full tournament simulations, since after a gambit, only the remainder of the tournament needs to be simulated. In the probabilistic model, each match result option (with a prefix tournament) was completed to a full tournament 200 times.

The choice of sample size 200 for the probabilistic model is based on initial experiments with sample size 50 and sample size 100. For these values, we observed very similar results that are in line with the results we present here. We chose sample size 200 since this was the maximum computational load our compute server could handle. Given our observations with different sample sizes, we expect that larger sample sizes yield very similar results but at an extraordinarily high computational cost. Thus, our choice seems to be sufficient for observing what can be observed.

Hence, a complete simulation including gambit possibilities is very expensive in terms of the required computing power, especially in the probabilistic model. This is why we only simulate 1000 tournaments with 5 rounds each.

The experiments were run on a computer server using version 20.04.1 of the Ubuntu operating system. It is powered by 48 Intel Xeon Gold 5118 CPUs running at 2.3 GHz and 62.4 GiB of RAM. With this server, simulating 1000 tournaments with 5 rounds each took approximately 100 seconds in the deterministic and 250 minutes in the probabilistic model.

\subsection{Presentation of the Results} We ran experiments in the deterministic and the probabilistic models. We discuss the effect of gambits in Swiss-system chess tournaments consisting of few or many rounds, and having a narrow or wide player strength range.

Data is presented in the form of \textit{violin plots} \cite{hintze1998violin} and via \textit{boxen plots}, which were invented by Heike et al.~\cite{heike2017letter}. For violin plots, kernel density estimation is used to show a smoothed probability density function of the underlying distribution. Additionally, similar to box plots, quartiles are shown by dashed lines. Boxen plots are enhanced box plots that show more quantiles. Unlike violin plots, they are suitable for discrete values, because all shown values are actual observations and there is no smoothing.

\section{Simulation Results}\label{sec:gambit_simulation_results}
In this section, we consider our four measures of gambit impact and elaborate on the obtained simulation results.

\subsection{Number of Gambit Possibilities}
\label{sec:number_of_gambit_possibilities}

First we ran experiments in our standard setting, changing only the number of rounds and letting the 32 players engage in more than 5 games each.

\begin{figure}[htb]
    \centering
    \includegraphics[width=0.8\linewidth]{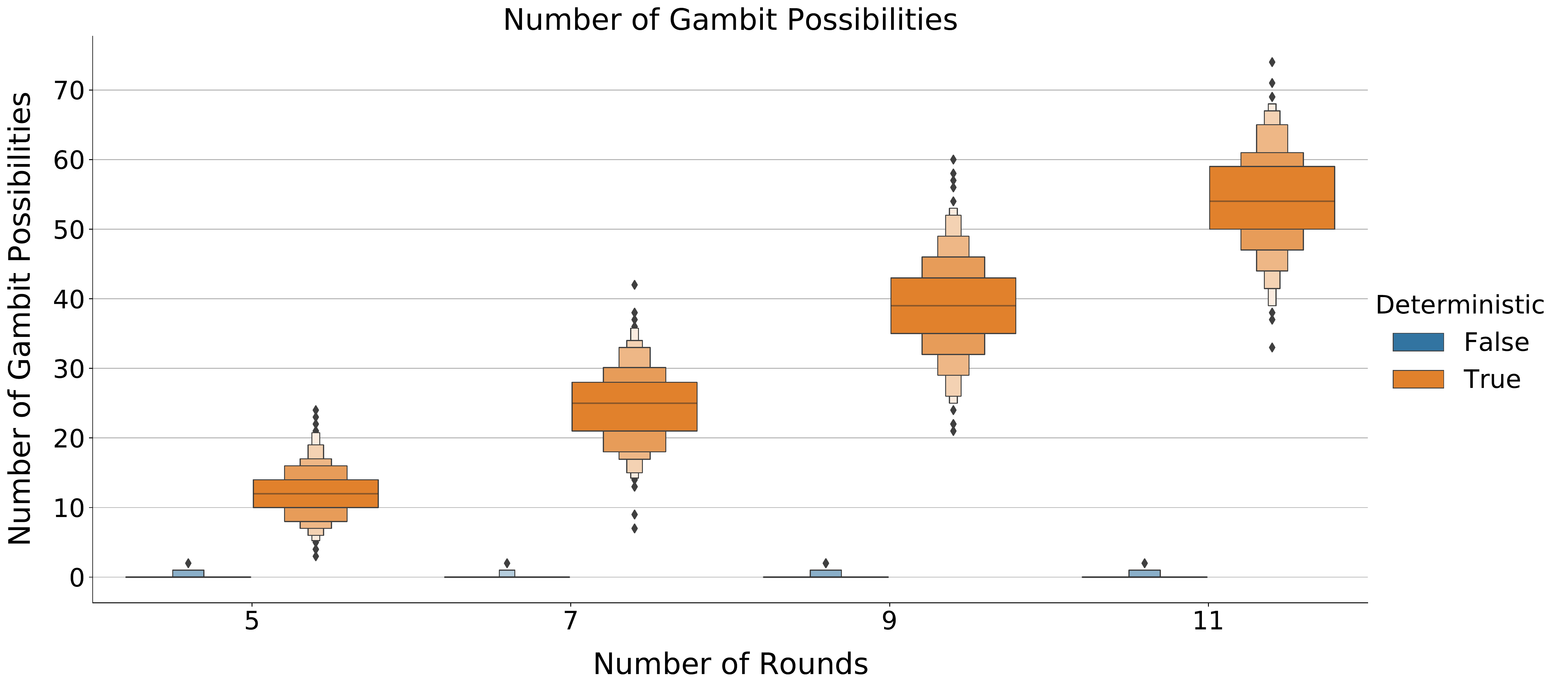}
    \caption{Number of gambit possibilities as a function of the number of rounds. Results for the deterministic model are shown in orange, while results for the probabilistic model are shown in blue.}
    \label{fig:gambits_vs_rounds}
\end{figure}

In the deterministic model, there are 12 gambit possibilities on average
in our standard setting, as Figure~\ref{fig:gambits_vs_rounds} shows. This number increases to 54 if there are 11 rounds. We can thus observe that the length of the tournament is a decisive factor in the number of possible gambits. This is to be expected, as the gambit player has more matches to capitalize on her gambit in longer tournaments.

\begin{figure}[htb]
    \centering
    \includegraphics[width=0.75\linewidth]{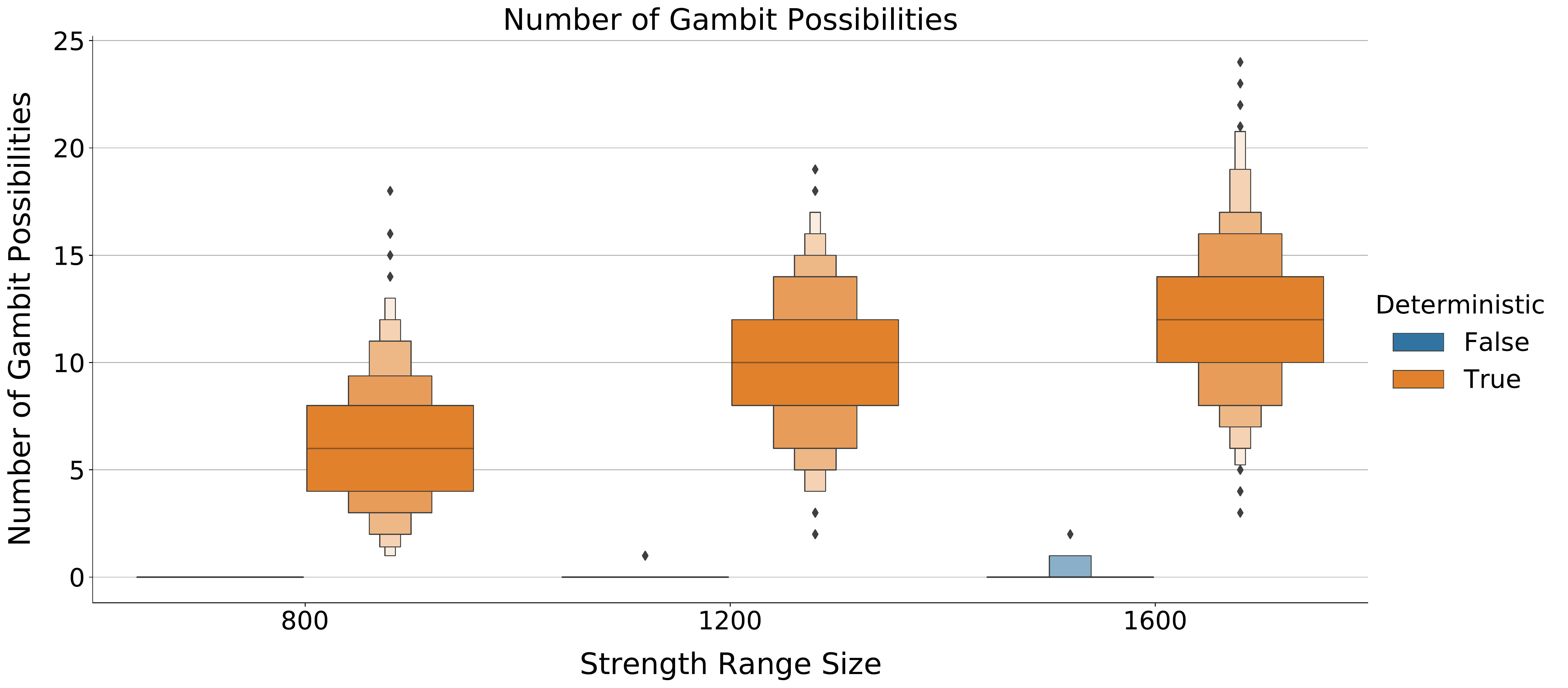}
    \caption{Number of gambit possibilities when player strengths are drawn from strength ranges of different size. Results for the deterministic model are shown in orange, results for the probabilistic model are shown in blue.}
    \label{fig:gambits_vs_strength_range_size}
\end{figure}

A less significant increase in the number of gambit possibilities is shown in Figure~\ref{fig:gambits_vs_strength_range_size}. As the strength range grows from 800 to 1600, the number of gambit possibilities also doubles. The center Elo was set to 1800 in all three settings, e.g., strength range size 800 means the interval [1400, 2200].

The reason for this correlation is not as straightforward as in the previous case. The increase in the number of gambit possibilities is driven by the fact that with a larger strength range size and given that we sample the player strengths uniformly in the strength range interval, the number of matches that result in a draw decreases. Thus, stronger players win more often which gives them two gambit options (choosing to draw or to lose) instead of only one (choosing to lose). Also, the lower number of draws also means that after a gambit, a strong player can earn more points and thus gambits have a higher chance to succeed.

In the probabilistic model, almost no gambit is possible, however, their sporadic occurrence becomes somewhat more frequent if the tournament consists of many rounds (see Figure~\ref{fig:gambits_vs_rounds}) or the strength range is large (see Figure~ \ref{fig:gambits_vs_strength_range_size}).

\begin{quote}
\textbf{Main Take-Away:} Under realistic conditions, i.e., in the probabilistic setting with a rather small strength range size, there are very few occasions when a gambit can improve a player's final rank. Moreover, the gambit player must be able to estimate match results very accurately in order to identify a gambit possibility.
\end{quote}

\subsection{Mean Rank Difference}
\label{sec:mean_rank_difference}

We measured the mean rank difference achieved by the gambit player. For example, a mean rank difference of -2 means that in an average gambit possibility, the gambit player will improve her final rank by two places.

\begin{figure}[ht]
    \centering
    \includegraphics[width=0.6\linewidth]{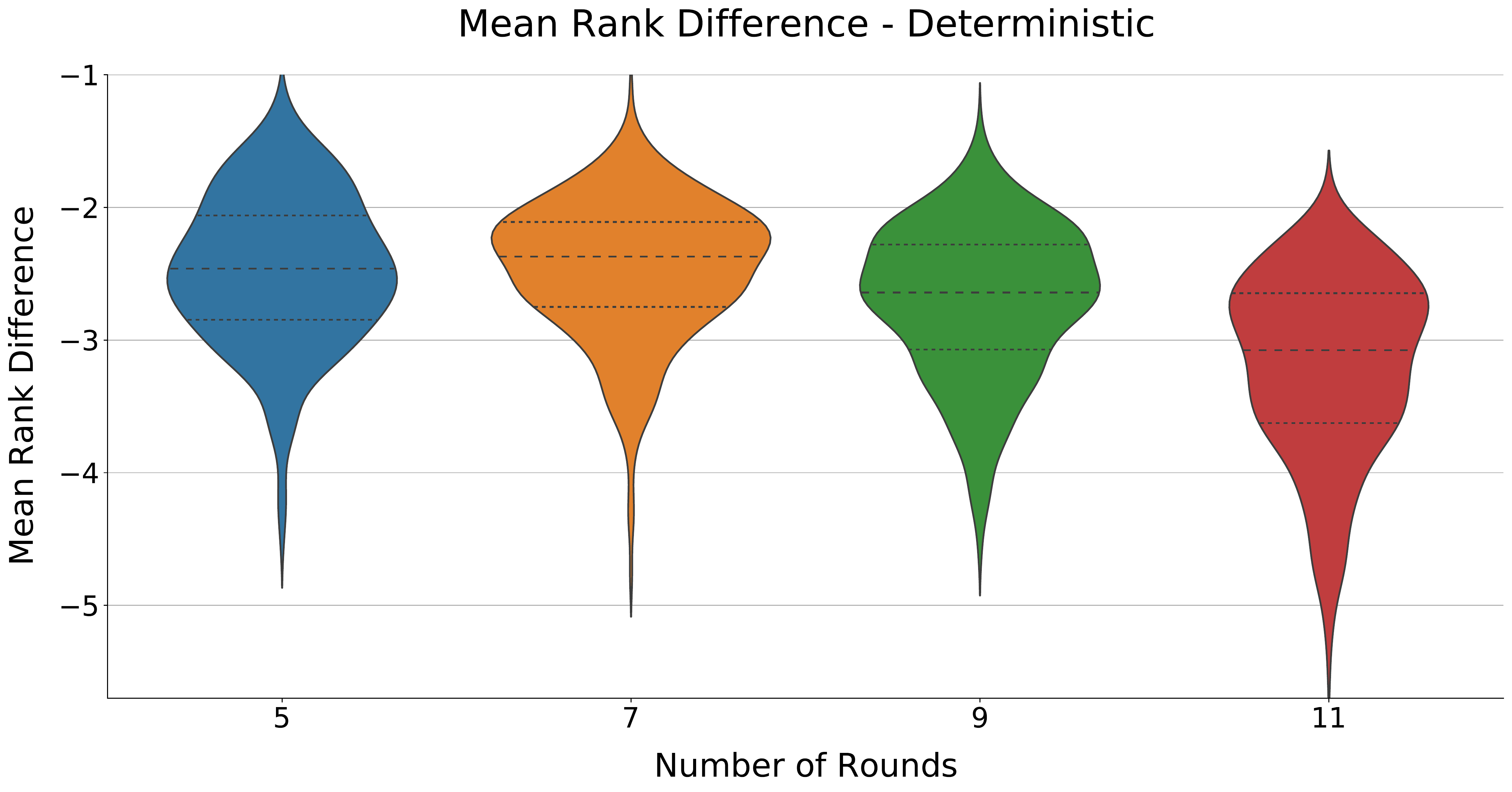}
    \caption{Mean rank difference for different numbers of rounds in the deterministic model.}
    \label{fig:mean_rank_diff_det_vs_number_of_rounds}
\end{figure}

In the deterministic model, as the number of rounds increases, we observe that gambits have a larger effect, but even for tournaments consisting of 11 rounds, an improvement of three places is to be expected, as shown by Figure~\ref{fig:mean_rank_diff_det_vs_number_of_rounds}.

\begin{figure}[ht]
    \centering
    \includegraphics[width=0.6\linewidth]{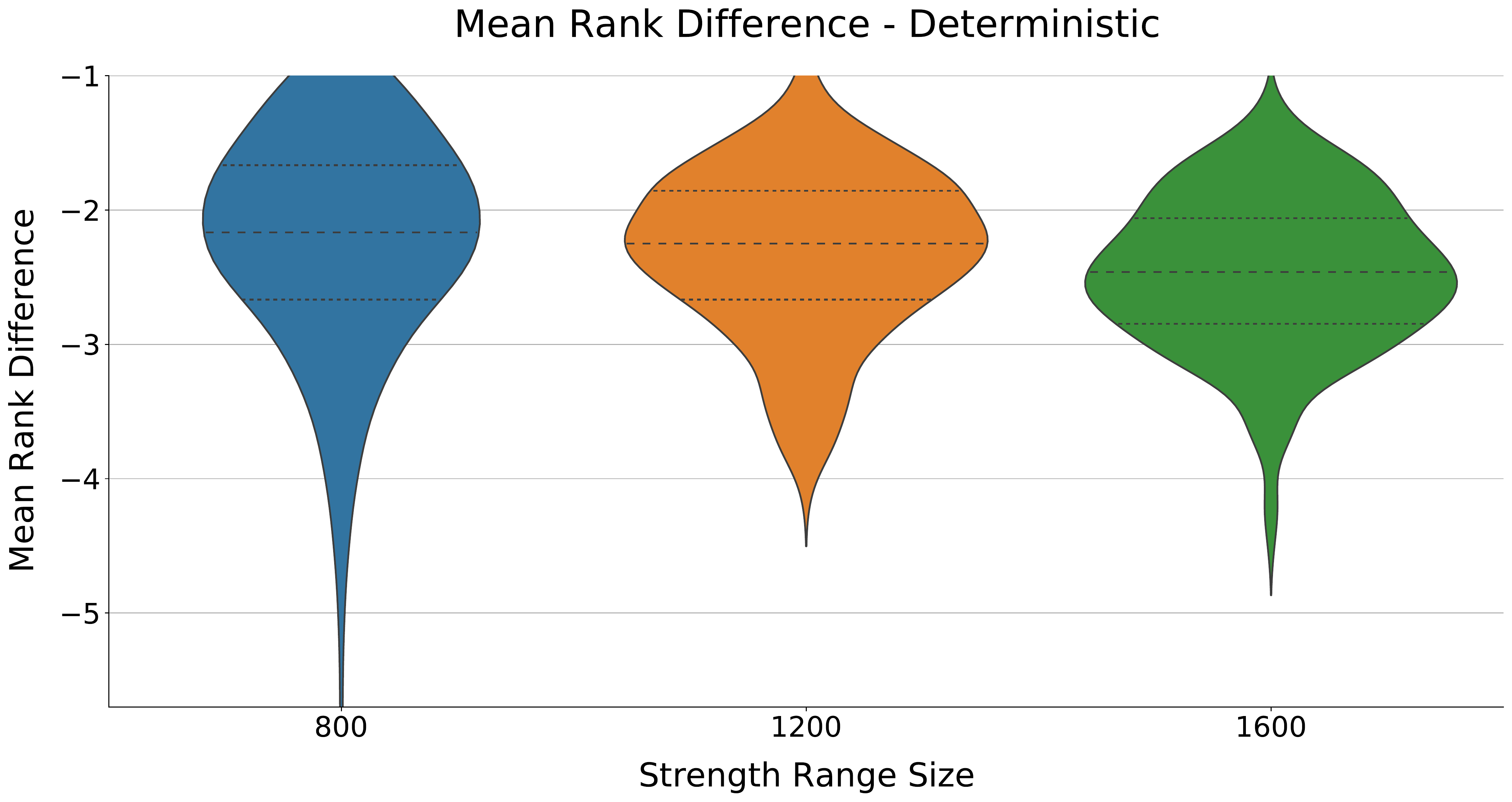}
    \caption{Mean rank difference for different strength range sizes in the deterministic model.}
    \label{fig:mean_rank_diff_det_vs_strength_range_size}
\end{figure}

Moreover, the larger the strength range size is, the more is to be gained through a gambit, as Figure~\ref{fig:mean_rank_diff_det_vs_strength_range_size} demonstrates.
The difference is rather small: In Swiss-system chess tournaments with a strength range of 800 Elo points, a bit less than half of the gambits lead to an improvement of at most two places in the final ranking, while it is a little less than a quarter of the gambits for the strength range of 1600 Elo points.

To summarize: in the deterministic model, performing a gambit is more beneficial if the tournament is longer or the players' strength level is very diverse.

\begin{figure}[ht]
    \centering
    \includegraphics[width=0.6\linewidth]{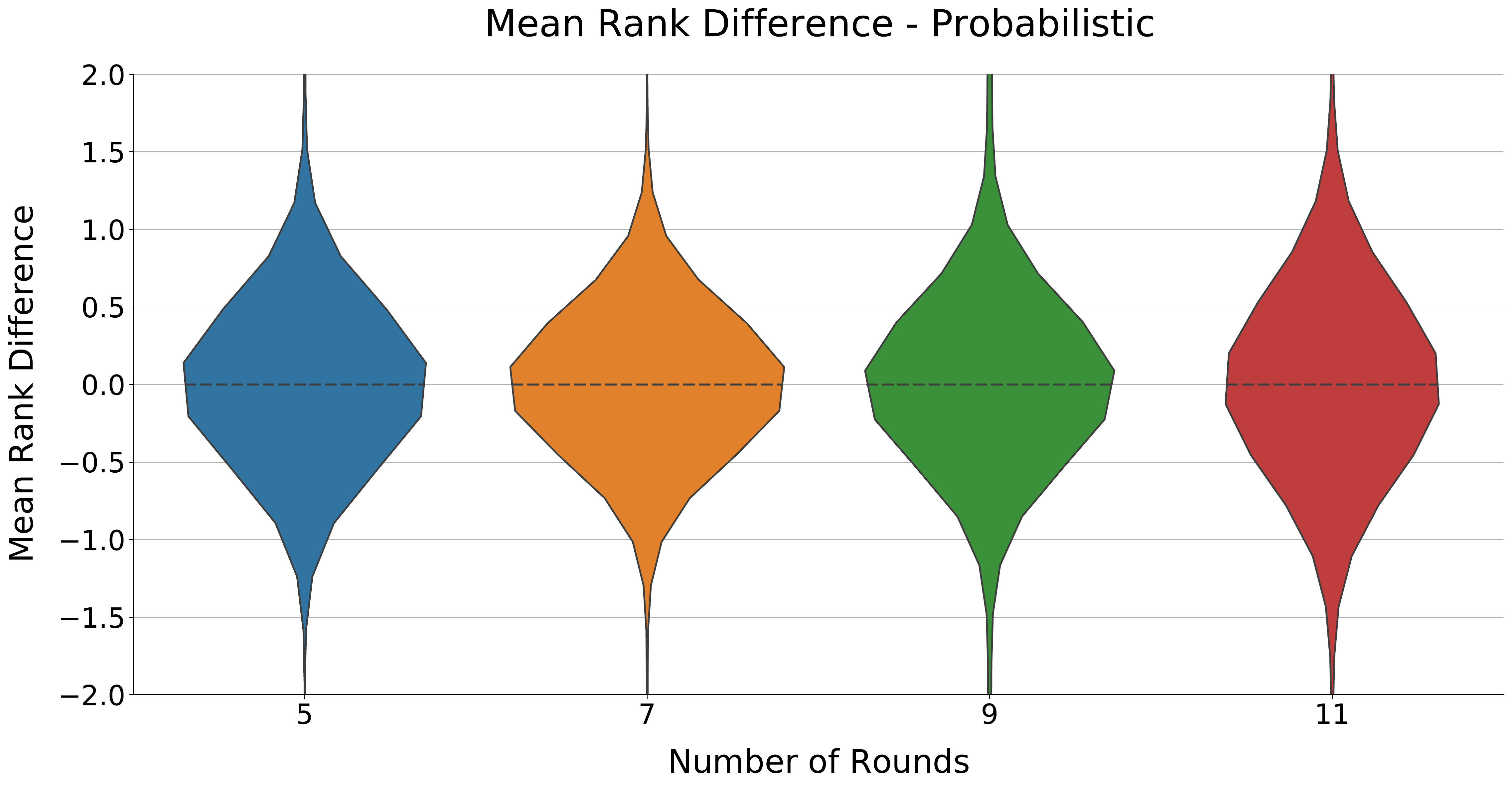}
    \caption{Mean rank difference for different numbers of rounds in the probabilistic model.}
    \label{fig:mean_rank_diff_prob_vs_number_of_rounds}
\end{figure}

\begin{figure}[ht]
    \centering
    \includegraphics[width=0.6\linewidth]{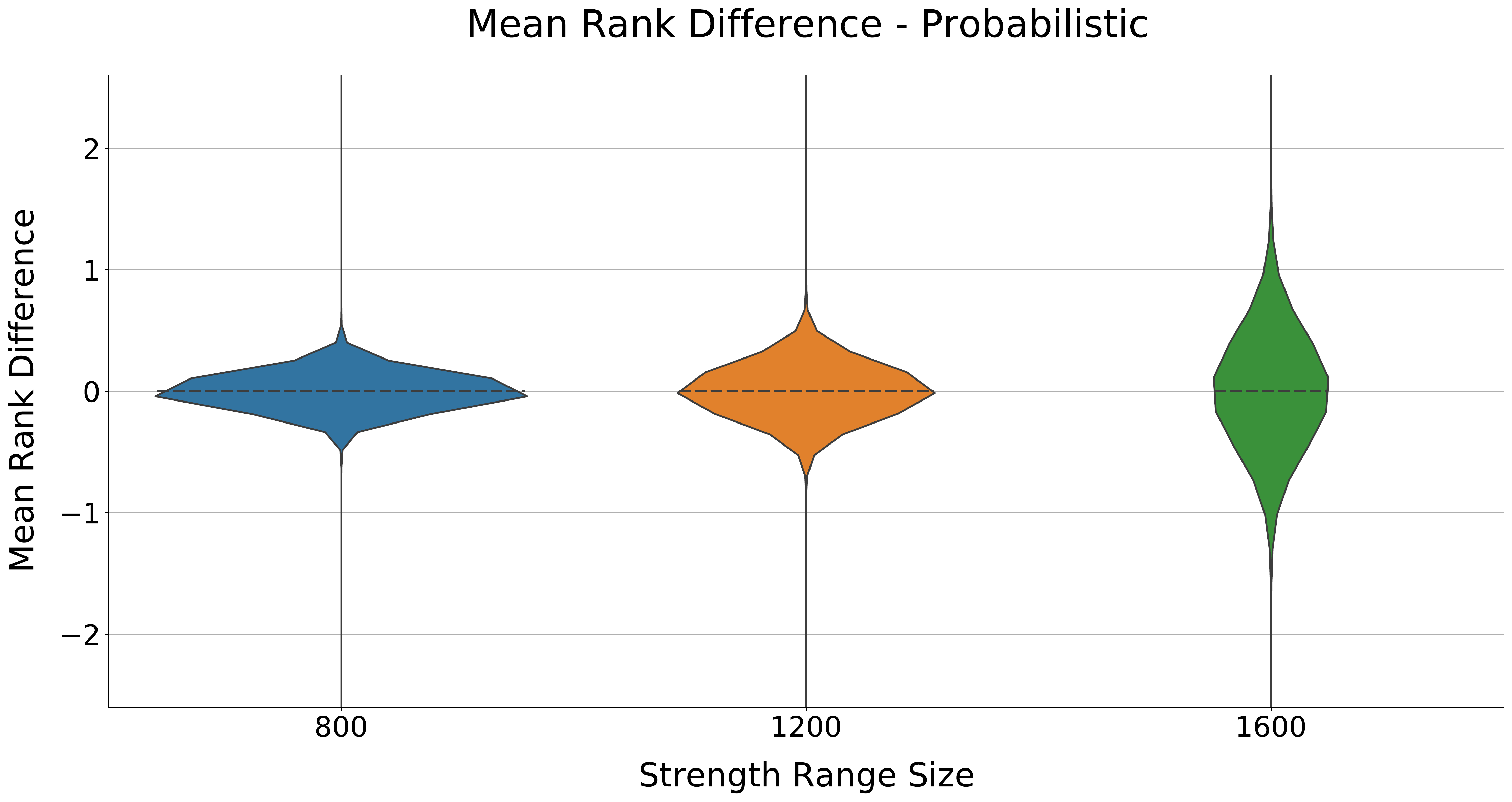}
    \caption{Mean rank difference for different strength range sizes and 7 rounds in the probabilistic model.}
    \label{fig:mean_rank_diff_prob_vs_strength_range_size}
\end{figure}

\paragraph{Mean Rank Difference in the Probabilistic Model} Mean rank differences in the probabilistic model are much closer to zero, as Figures~\ref{fig:mean_rank_diff_prob_vs_number_of_rounds} and \ref{fig:mean_rank_diff_prob_vs_strength_range_size} show. Note that due to the probabilistic nature of the model, gambits that work in expectation still frequently lead to a positive rank difference. Ending up at rank difference close to 0 on average means that our chosen heuristic performs well, considering that the gambit player gives up a safe (half) point when performing a gambit.

Figure~\ref{fig:mean_rank_diff_prob_vs_number_of_rounds} shows that the number of rounds does not seem to influence the mean rank difference much in the probabilistic model: the variance grows only very slightly with the number of rounds.

The strength range size slightly influences the mean rank difference, as Figure~\ref{fig:mean_rank_diff_prob_vs_strength_range_size} demonstrates. As the strength range size grows, the variance grows as well, suggesting that performing a gambit is more risky for a diverse player set.

Notice that giving up points without using any heuristic would lead to a strong positive rank difference: players would lose several ranks. The fact that the mean rank difference is approximately zero in the probabilistic model shows the main message of this subsection, namely that the $p$-value heuristic can identify gambit possibilities with reasonable accuracy. These gambits are slightly riskier if the tournament has many rounds (Figure~\ref{fig:mean_rank_diff_prob_vs_number_of_rounds}) or if the strength range is large (Figure~\ref{fig:mean_rank_diff_prob_vs_strength_range_size}).

\begin{quote}
\textbf{Main Take-Away:}
If all match results can be accurately predicted, i.e., in the deterministic model, gambits are slightly more beneficial the more rounds are played and the larger the strength range of the players is. This situation drastically changes in the probabilistic model. There, gambits detected by our $p$-value heuristic yield a mean rank difference of zero in expectation, i.e., they are not profitable. Note that other natural gambit detection heuristics fare much worse in this regard.
\end{quote}

\subsection{Total Rank Difference}\label{sec:total_rank_difference}

The total rank difference incorporates the number of gambit possibilities and the mean rank difference, thus it provides a general indicator for how much gambits are incentivized. A total rank difference close to zero indicates that gambit possibilities are mostly prevented.

\begin{figure}[ht]
    \centering
    \includegraphics[width=0.6\linewidth]{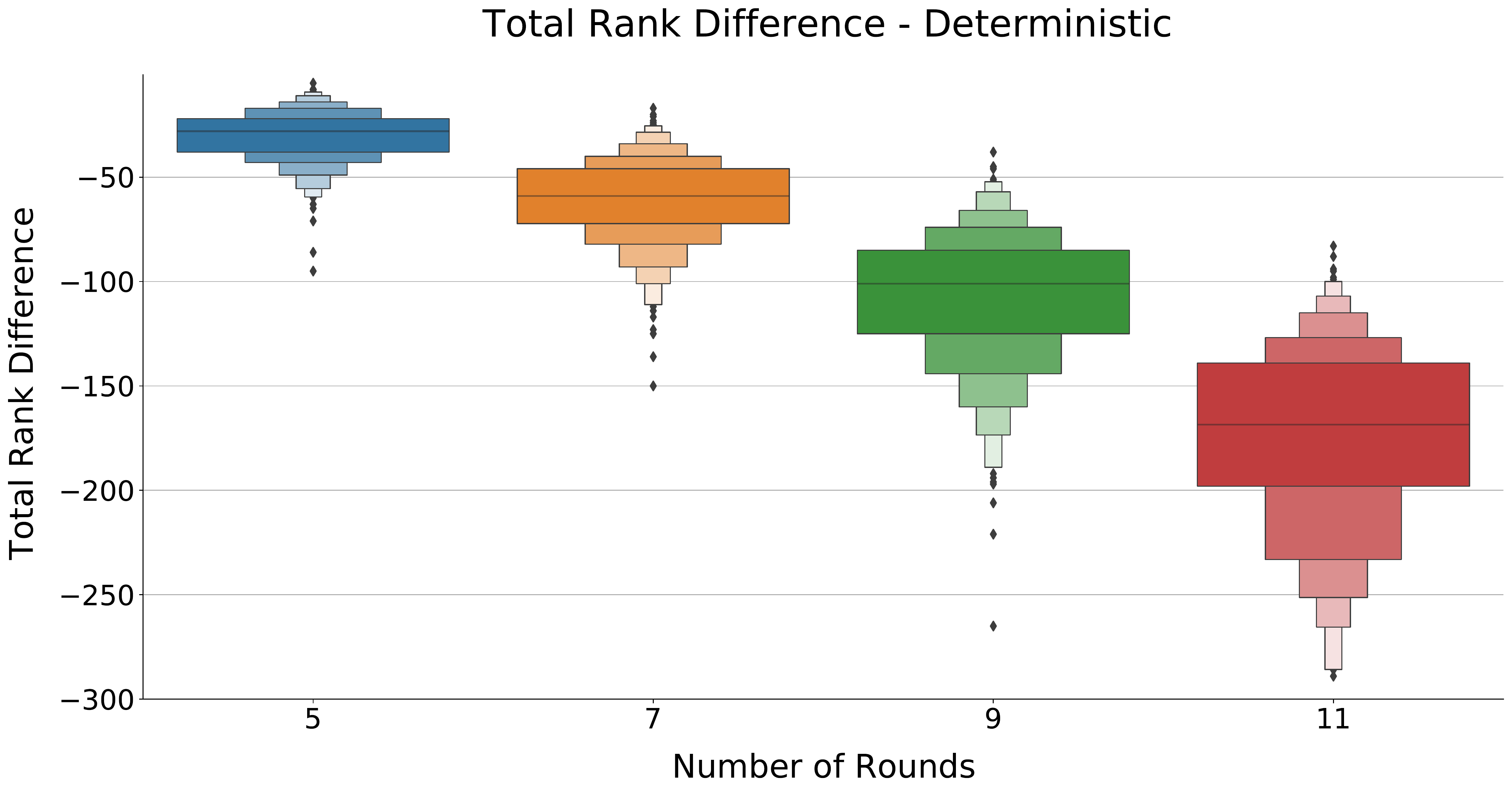}
    \caption{Total rank difference for different number of rounds in the deterministic model.}
    \label{fig:total_rank_diff_det_vs_number_of_rounds}
\end{figure}

\begin{figure}[ht]
    \centering
    \includegraphics[width=0.6\linewidth]{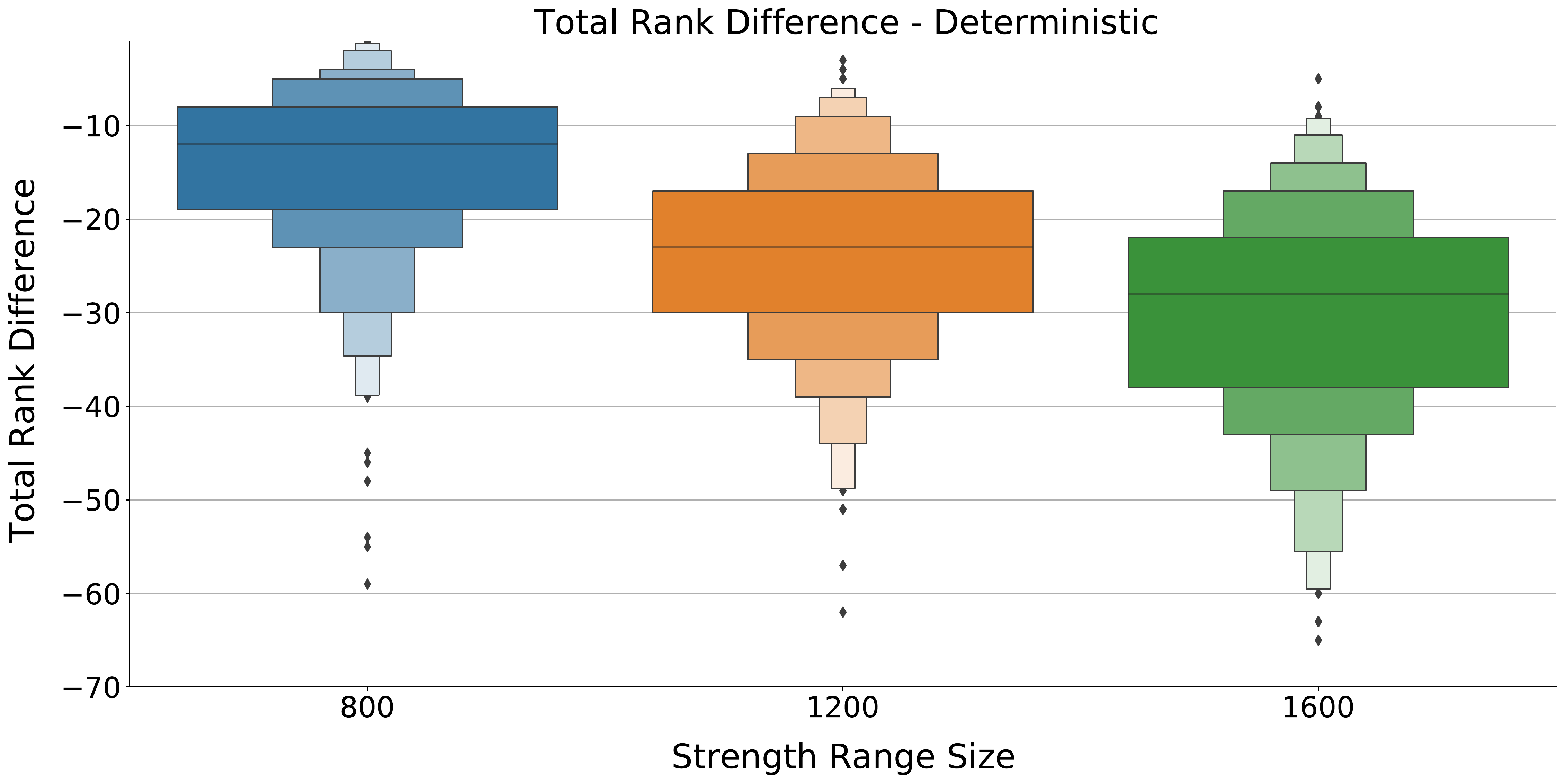}
    \caption{Total rank difference for different strength range sizes in the deterministic model.}
    \label{fig:total_rank_diff_det_vs_strength_range_size}
\end{figure}

Figure~\ref{fig:total_rank_diff_det_vs_number_of_rounds} shows that in the deterministic model and standard setting the total rank difference sharply increases with the number of rounds. For example, for 11 rounds we obtain on average a total rank difference of 170, compared to 28 for 5 rounds. We also see in Figure~\ref{fig:total_rank_diff_det_vs_strength_range_size} that the total rank difference increases with the strength range size. Both observations are in line with our results from Sections~\ref{sec:number_of_gambit_possibilities} and~\ref{sec:mean_rank_difference}.

\paragraph{Total Rank Difference in the Probabilistic Model}
As we have shown in Section~\ref{sec:number_of_gambit_possibilities}, gambits occur very rarely in the probabilistic model. Therefore, drawing consequences from the few data we could collect on them is somewhat hard.

\begin{figure}[ht]
    \centering
    \includegraphics[width=0.6\linewidth]{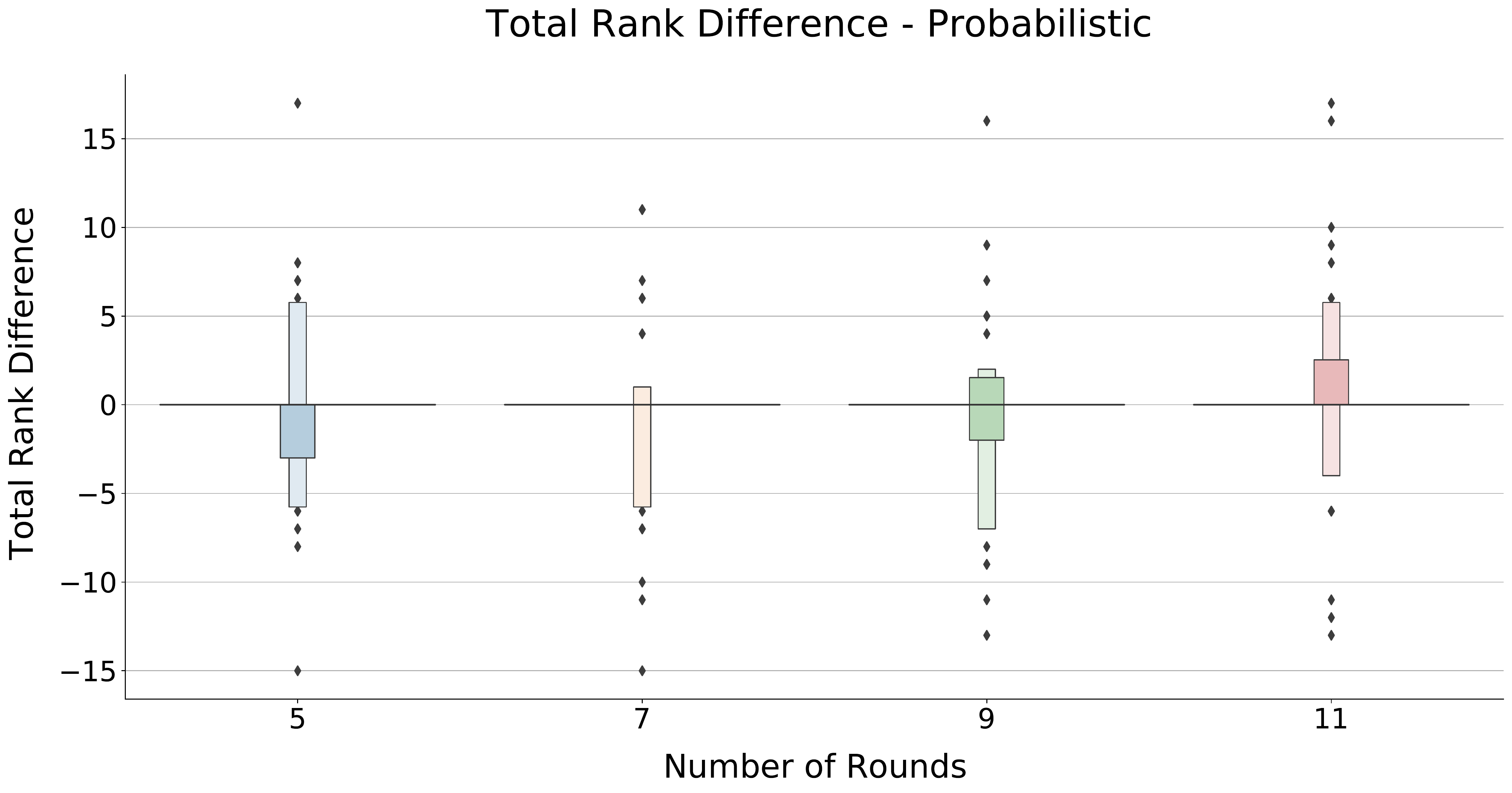}
    \caption{Total rank difference for different number of rounds in the probabilistic model.}
    \label{fig:total_rank_diff_prob_vs_number_of_rounds}
\end{figure}

In the probabilistic model, players can only improve their rank marginally by performing a gambit. Figure~\ref{fig:total_rank_diff_prob_vs_number_of_rounds} displays an unexpected correlation between the total rank difference and the number of rounds in the probabilistic model. As the number of rounds grow, there is very slightly less to win.
The reason for this unexpected behavior might be that with an increasing number of rounds, it becomes less likely to correctly predict the remaining tournament rounds. This then increases the risk of losing ranks by performing a gambit, yielding a positive total rank difference.

\begin{figure}[ht]
    \centering
    \includegraphics[width=0.6\linewidth]{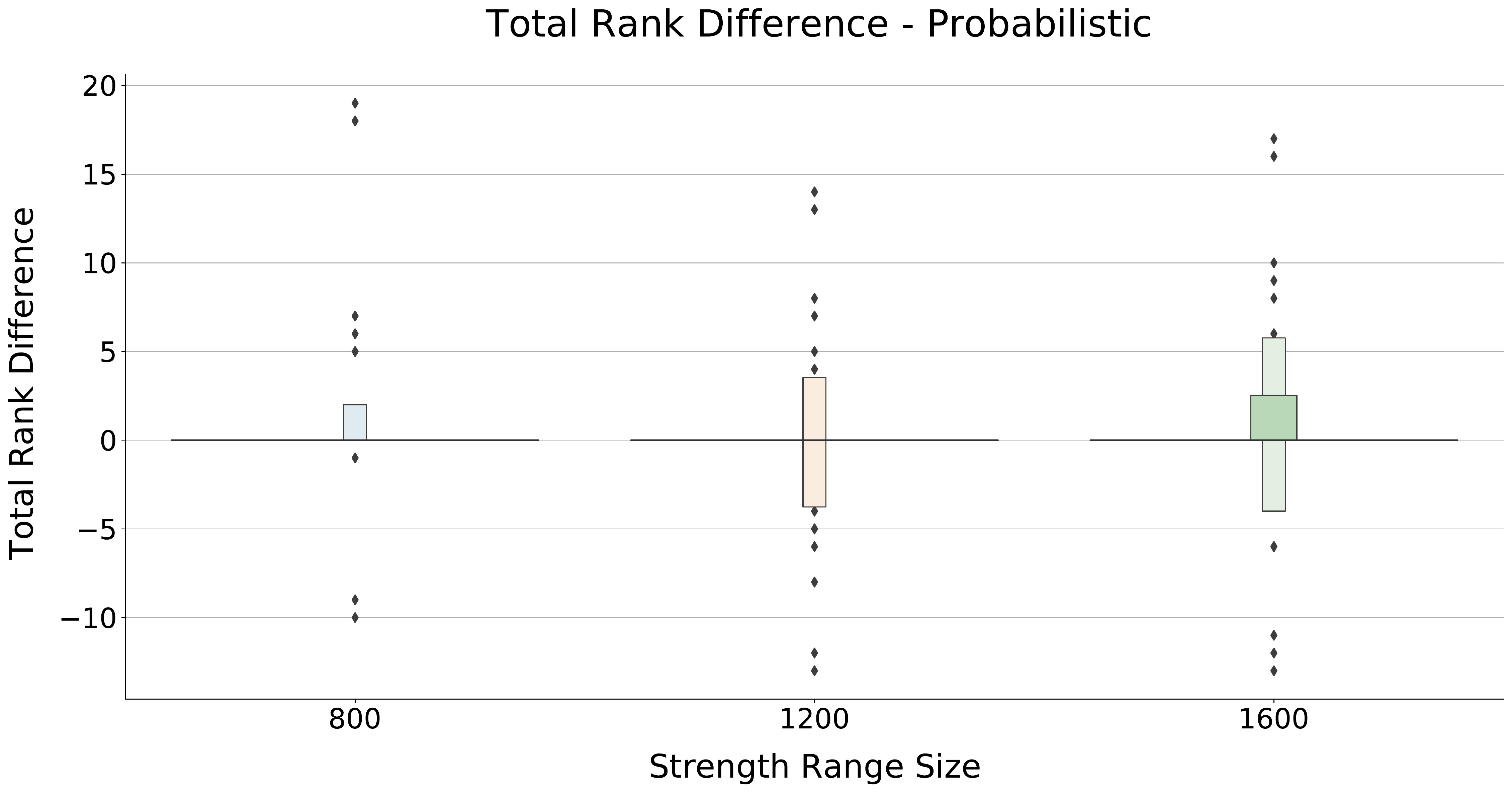}
    \caption{Total rank difference for 11 rounds and different strength range sizes in the probabilistic model.}
    \label{fig:total_rank_diff_prob_vs_strength_range_size}
\end{figure}

In order to compare different strength range sizes for the probabilistic model, we had to deviate from our standard setting and set the number of rounds to 11, because for shorter tournaments, the strength range size did not make a noticeable difference. Figure~\ref{fig:total_rank_diff_prob_vs_strength_range_size} shows that for 11 rounds, a more diverse player set leads to a somewhat larger variance in the total rank difference. The same is shown in Figure~\ref{fig:total_rank_diff_det_vs_strength_range_size}, which is the corresponding plot in the deterministic model.

\begin{quote}
\textbf{Main Take-Away:}
The results regarding the total rank difference are in line with the results regarding the mean rank difference. In the deterministic model, gambits become slightly more beneficial with increasing number of rounds or with larger player strength range. In the probabilistic model, gambits are not beneficial in expectation.
\end{quote}

\subsection{Ranking Quality}\label{sec:impact_of_gambits_on_ranking_quality}

We investigate the ranking quality difference between tournaments with and without gambits. In each of the 1000 simulated tournaments, the value without gambit is the Kendall $\tau$ of the simulation without gambits, while the value with gambit is the mean of the Kendall $\tau$ values of all gambit possibility simulations.

\begin{figure}[h!]
    \centering
    \includegraphics[width=0.8\linewidth]{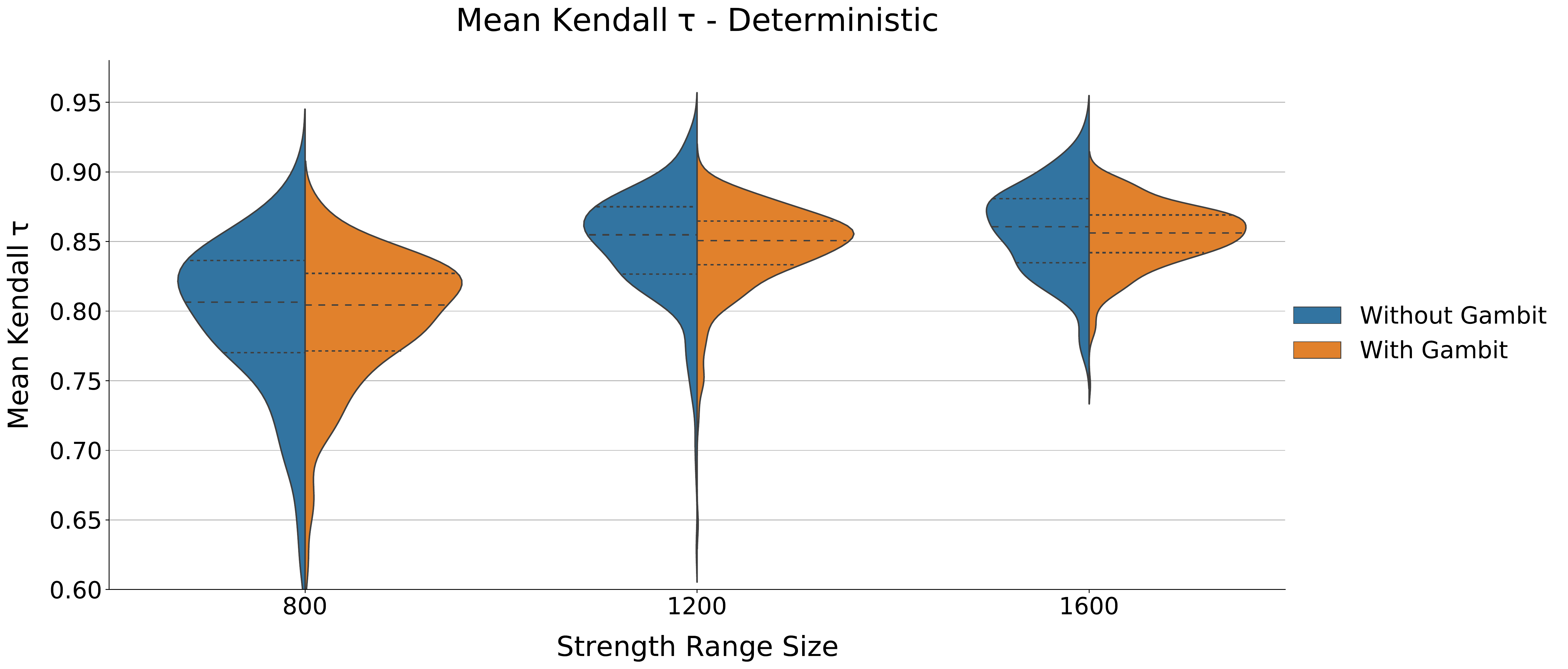}
    \caption{Ranking quality with and without gambits for different strength ranges in the deterministic model.}
    \label{fig:ranking_quality_with_vs_without_gambit_det_strength}
\end{figure}

\begin{figure}[htb]
    \centering
    \includegraphics[width=0.8\linewidth]{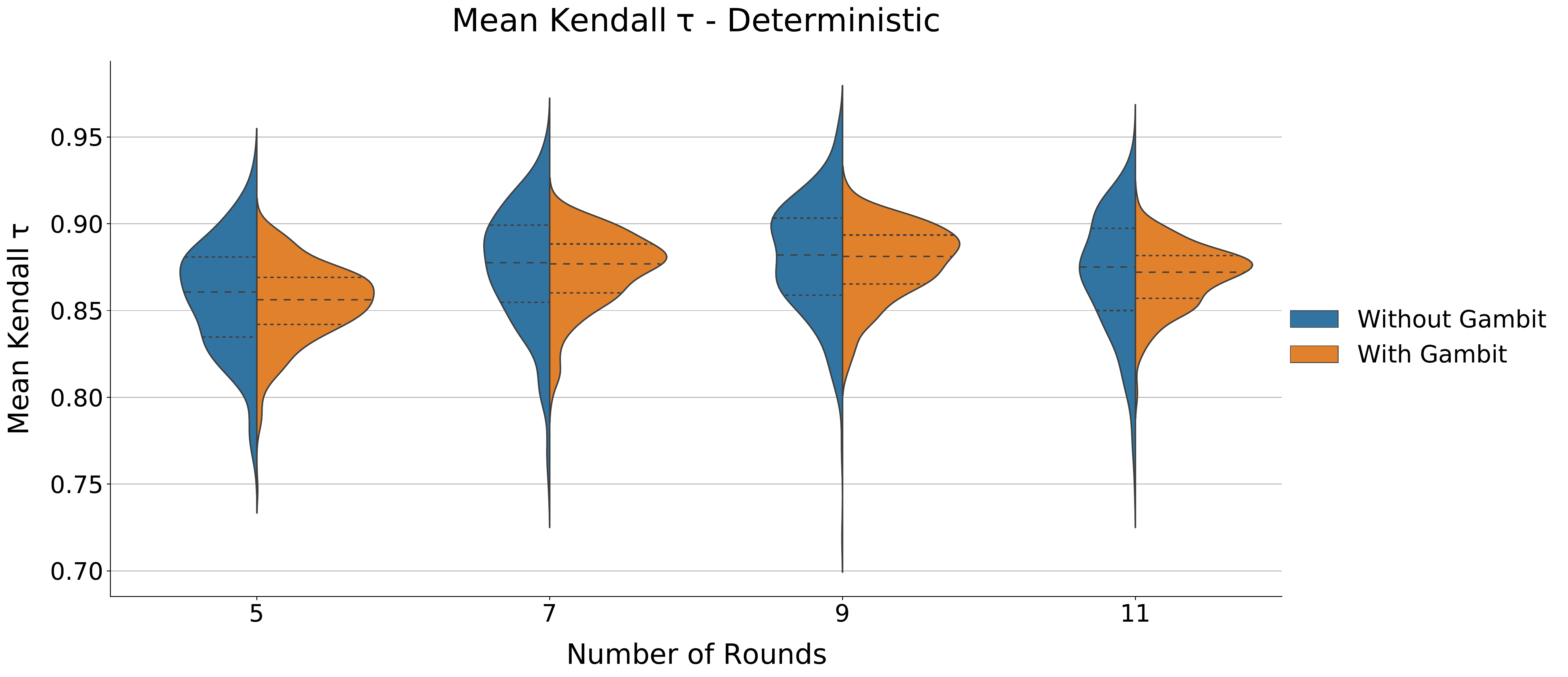}
    \caption{Obtained ranking quality for different numbers of rounds in the deterministic model. }
    \label{fig:ranking_quality_with_vs_without_gambit_det_rounds}
\end{figure}

Figures~\ref{fig:ranking_quality_with_vs_without_gambit_det_strength} and \ref{fig:ranking_quality_with_vs_without_gambit_det_rounds} show that the impact of gambits on the ranking quality is low in general in the deterministic model. However, gambits spoil the ranking quality at some level, especially if the strength range is large or the tournament is longer. Also, note that in our plots the results for the variant with gambits seems stronger concentrated around the mean Kendall $\tau$ values than the results without gambit. The reason for this is that without gambit the Kendall $\tau$ value of a single simulation is shown while with gambit, the mean Kendall $\tau$ values of all gambit possibility simulations is plotted. This naturally concentrates the values around the mean.

\begin{figure}[htb]
    \centering
    \includegraphics[width=0.8\linewidth]{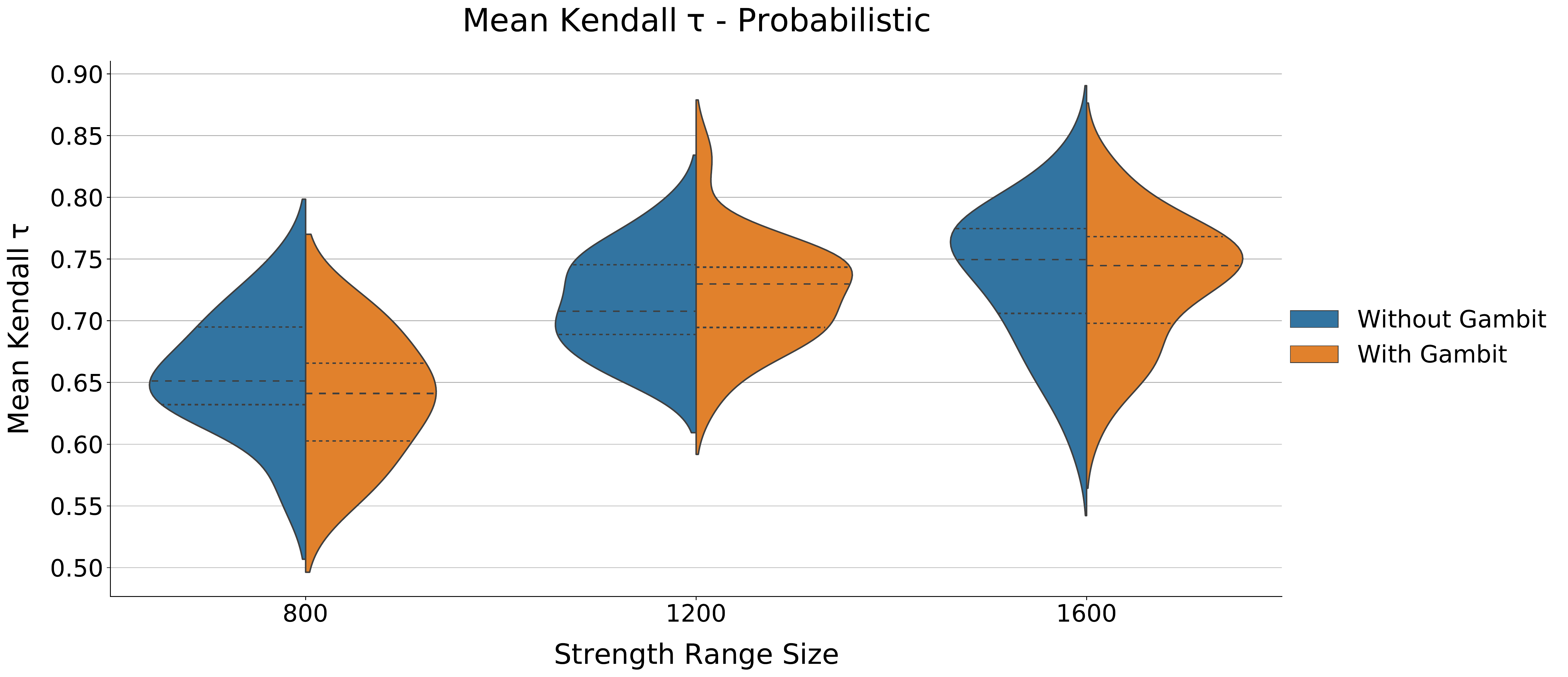}
    \caption{Obtained ranking quality for different strength range sizes in the probabilistic model.}
    \label{fig:ranking_quality_with_vs_without_gambit_prob_strength}
\end{figure}

\begin{figure}[htb]
    \centering
    \includegraphics[width=0.8\linewidth]{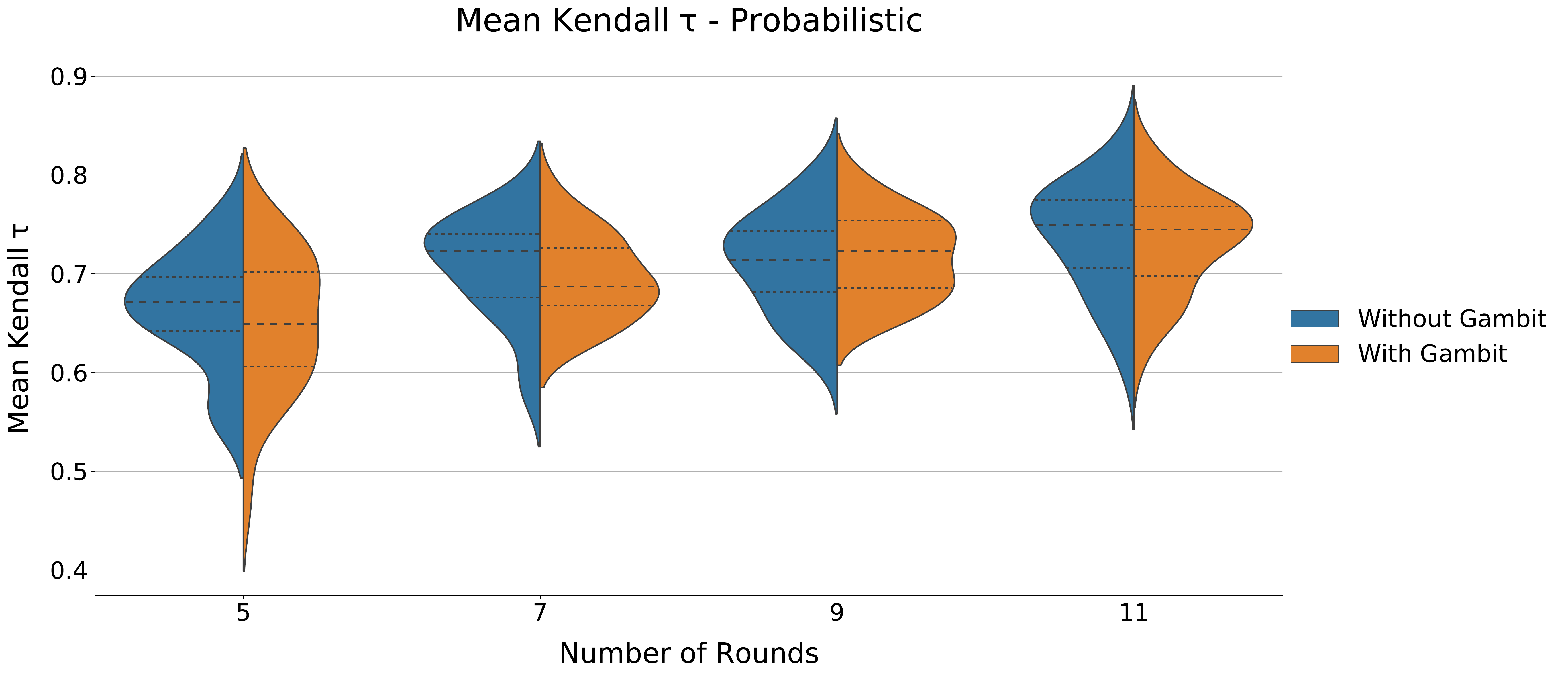}
    \caption{Obtained ranking quality for different number of rounds in the probabilistic model after 11 rounds. }
    \label{fig:ranking_quality_with_vs_without_gambit_prob_rounds}
\end{figure}

\paragraph{Ranking Quality in the Probabilistic Model}
Figures~\ref{fig:ranking_quality_with_vs_without_gambit_prob_strength} and~\ref{fig:ranking_quality_with_vs_without_gambit_prob_rounds} show that the impact of gambits on ranking quality is low in general in the probabilistic model as well.
Gambits spoil the ranking quality in most, but not all cases. Just as in Section~\ref{sec:total_rank_difference}, we had to deviate from our standard setting and set the number of rounds to 11 for the Figure~\ref{fig:ranking_quality_with_vs_without_gambit_prob_strength}, because no noticeable difference was to be seen for shorter tournaments.

\begin{quote}
\textbf{Main Take-Away:} The impact of gambits on the ranking quality is low in both the deterministic and the probabilistic model.
\end{quote}

\section{Conclusion} 

We have shown that even though the Swiss Gambit is possible in theory, identifying the match in which to perform it is extremely challenging and even a beneficial gambit comes with a relatively low rank improvement.

As we are the first ones to study the Swiss Gambit from a scientific point of view, our work raises various open questions.
\begin{itemize}
    \item Our simulations can be run on other types of tournaments, such as longer events for professional players, which would require more rounds and a smaller strength range. Clearly, gambling the Swiss-system in other sports or games could also be analyzed, if sufficient data is available.
    \item We identified an effective gambit heuristic in the probabilistic model, but there might be even smarter heuristics---possibly depending on the tournament type. 
    \item Based on the players' Elo scores, one might attempt to identify Swiss Gambits---or very fortunate unexpectedly poor match results---performed at real tournaments; these are unexpected losses or draws from strong players who then reached a higher rank than they would have otherwise.
    \item With our tournament simulation we estimate the importance of a chosen match result. What is the expected final rank difference of the upcoming match is lost versus if it is won?
    \item Our investigation can be tailored to answer questions about cheating during a match~\citep{Kee22} or bribing the opponent~\citep{Win22}, both being actively discussed in the chess community, especially since computers significantly changed the way chess is played~\citep{Won22}. How much is there to win in expectation if a chosen match result is turned positive?
    \item Finally, a strategyproof modification of the Swiss system that prevents Swiss Gambits would be highly valuable. Although our results indicate that gambits might not be a problem in practice, the online discussions and newspaper articles mentioned in the introduction demonstrate that the sheer (theoretical) possibility of a Swiss Gambit already irks the community.
\end{itemize}

\section*{Acknowledgments}
\'{A}gnes Cseh's work was supported by OTKA grant K128611, the J\'anos Bolyai Research Fellowship, and \'UNKP-22-5-ELTE-1166 grant.

\bibliographystyle{ACM-Reference-Format}
\bibliography{references}

\newpage
\appendix

\section{The Dutch Pairing System}\label{app:dutch}

Players are ranked by their current tournament score in each round; players with equal score are grouped into \textit{score groups}. Within each score group, the players are ranked by their Elo rank. The goal is to pair players within their score group in each round. If a complete pairing is not possible within a score group, then one or more players are moved to another score group with a similar score.

In each round, a chosen \textit{pairing system} allocates each player an opponent from the same or a similar score group. Three main pairing systems are defined for chess tournaments. Table~\ref{tab:pairing_systems_example_pairing} shows an example pairing for each of them.
\begin{itemize}
\item \textbf{Dutch:}
Each score group is cut into an upper and a lower half. The upper half is then paired against the lower half so that the $i$th ranked player in the upper half plays against the $i$th ranked player in the lower half. Dutch is the de facto standard for major chess tournaments.
\item \textbf{Burstein:}
For each score group, the highest ranked unpaired player is paired against the lowest ranked unpaired player repeatedly until all players are paired.
\item \textbf{Monrad:}
In ascending rank order each unpaired player in a score group is paired against the next highest ranked player in that score group.
\end{itemize}

\begin{table}[ht]
\centering
\begin{tabular}{ccccccc}
Dutch   &&& Burstein    &&& Monrad\\
1--5     &&& 1--8         &&& 1--2\\
2--6     &&& 2--7         &&& 3--4\\
3--7     &&& 3--6         &&& 5--6\\
4--8     &&& 4--5         &&& 7--8\\
\end{tabular}
\caption{Example pairing for each pairing system in a score group of 8 players. Players are referenced by rank within the score group, i.e., player 1 has the highest Elo rank.}
\label{tab:pairing_systems_example_pairing}
\end{table}

Besides pairing players within their score group, the pairing engine also needs to keep the color assignment balanced. The FIDE Handbook~\cite[Chapter C.04.1]{fide2020handbook} states that `For each player the difference between the number of black and the number of white games shall not be greater than 2 or less than -2.' This criterion may only be relaxed in the last round. Also, a ban on a color that is assigned to a player three times consecutively, and further milder criteria are phrased to ensure a color assignment as close to an alternating white-black sequence as possible \cite[Chapters C.04.3.A.6 and C.04.3.C]{fide2020handbook}.

\end{document}